\documentclass[superscriptaddress,amsmath,amssymb,aps,prm,twocolumn,nofootinbib,longbibliography]{revtex4-2}
\usepackage{graphicx}         

\usepackage{textcmds}
\usepackage{dcolumn}          
\usepackage{bm}               
\usepackage{upgreek}
\usepackage{booktabs} 				
\usepackage{tabularx}
\usepackage{array}
\usepackage[colorlinks=true,urlcolor=blue,breaklinks=true]{hyperref}
\usepackage{xcolor}
\usepackage[a4paper,left=2cm,right=2cm]{geometry}
\usepackage{orcidlink}
\usepackage{gensymb}
\usepackage{makecell}

\usepackage{titlesec}

\titleformat{\section}
  {\centering\small\bfseries\uppercase} 
  {\thesection.}                            
  {1em}{}                                                            

\begin{document}

\title{Controlling the magneto-transport properties of magnetic topological insulator thin films from Cr$_x$(Bi$_y$\,Sb$_{1-y}$)$_{2-x}$Te$_3$ via molecular beam epitaxy}

\author{Jan Karthein\,\orcidlink{0009-0009-6887-8016}}
\email{j.karthein@fz-juelich.de.de}
\affiliation{Peter Gr\"unberg Institut (PGI-9), Forschungszentrum J\"ulich, 52425 J\"ulich, Germany}
\affiliation{JARA-Fundamentals of Future Information Technology, J\"ulich-Aachen Research Alliance, Forschungszentrum J\"ulich and RWTH Aachen University, Germany}

\author{Jonas Buchhorn\,\orcidlink{0009-0008-5437-7010}}
\affiliation{Peter Gr\"unberg Institut (PGI-9), Forschungszentrum J\"ulich, 52425 J\"ulich, Germany}
\affiliation{JARA-Fundamentals of Future Information Technology, J\"ulich-Aachen Research Alliance, Forschungszentrum J\"ulich and RWTH Aachen University, Germany}

\author{Kaycee Underwood\,
\orcidlink{0000-0003-1350-6222}}
\thanks{Current affiliation: Donostia International Physics Center (DIPC), 20018 Donostia-San Sebastián, Spain}
\affiliation{Peter Gr\"unberg Institut (PGI-9), Forschungszentrum J\"ulich, 52425 J\"ulich, Germany}
\affiliation{JARA-Fundamentals of Future Information Technology, J\"ulich-Aachen Research Alliance, Forschungszentrum J\"ulich and RWTH Aachen University, Germany}

\author{Abdur Rehman Jalil\,\orcidlink{0000-0003-1869-2466}}
\affiliation{Peter Gr\"unberg Institut (PGI-9), Forschungszentrum J\"ulich, 52425 J\"ulich, Germany}
\affiliation{Institute of Experimental Physics III, University of Würzburg, 97070 Würzburg, Germany}

\author{Max Vaßen-Carl\,\orcidlink{0000-0003-0895-7176}}
\affiliation{Peter Gr\"unberg Institut (PGI-9), Forschungszentrum J\"ulich, 52425 J\"ulich, Germany}
\affiliation{JARA-Fundamentals of Future Information Technology, J\"ulich-Aachen Research Alliance, Forschungszentrum J\"ulich and RWTH Aachen University, Germany}

\author{Peter Sch\"uffelgen\,\orcidlink{0000-0001-7977-7848}}
\affiliation{Peter Gr\"unberg Institut (PGI-9), Forschungszentrum J\"ulich, 52425 J\"ulich, Germany}
\affiliation{JARA-Fundamentals of Future Information Technology, J\"ulich-Aachen Research Alliance, Forschungszentrum J\"ulich and RWTH Aachen University, Germany}

\author{Detlev Gr\"utzmacher\,\orcidlink{0000-0001-6290-9672}}
\affiliation{Peter Gr\"unberg Institut (PGI-9), Forschungszentrum J\"ulich, 52425 J\"ulich, Germany}
\affiliation{JARA-Fundamentals of Future Information Technology, J\"ulich-Aachen Research Alliance, Forschungszentrum J\"ulich and RWTH Aachen University, Germany}

\author{Thomas Sch\"apers\,\orcidlink{0000-0001-7861-5003}}
\email{th.schaepers@fz-juelich.de}
\affiliation{Peter Gr\"unberg Institut (PGI-9), Forschungszentrum J\"ulich, 52425 J\"ulich, Germany}
\affiliation{JARA-Fundamentals of Future Information Technology, J\"ulich-Aachen Research Alliance, Forschungszentrum J\"ulich and RWTH Aachen University, Germany}

\hyphenation{}
\date{\today}

\begin{abstract}
In this work we present a systematic in-depth study of how we can alter the magneto-transport properties of magnetic topological insulator thin films by tuning the parameters of the molecular beam epitaxy. First, we show how a varying substrate temperature changes the surface morphology and when chosen properly leads to a high crystal quality. Next, the effect of the chromium concentration on the film roughness and crystal quality is investigated. Finally, both the substrate temperature and the chromium concentration are investigated with respect to their effect on the magneto-transport properties of the magnetic topological insulator thin films. It becomes apparent that the substrate temperature and the chromium concentration can be used to tune the Fermi level of the film which allows to make the material intrinsically charge neutral. A very low chromium concentration furthermore allows to tune the magnetic topological insulator into a regime where strong superconducting correlations can be expected when combining the material with a superconductor.  
\end{abstract}

\maketitle

\section{Introduction}

Recently, magnetic topological insulators (MTIs) have been proposed as a promising material for the realization of scalable topological quantum computation based on braiding of Majorana fermions~\cite{chen2018quasi,qi2010chiral}. MTIs are formed by incorporating magnetic adatoms, such as chromium or vanadium, into a topological insulator lattice~\cite{chang2013thin,zhang2017ferromagnetism}. Compared to regular topological insulators, the intrinsic magnetization gives rise to topologically protected edge states~\cite{yu2010quantized} that make the use of external magnetic fields for qubit applications obsolete~\cite{tokura2019magnetic}.  As a hallmark of the presence of edge channel transport, the quantum anomalous Hall effect (QAHE) has been measured, which is characterized by a quantized Hall signal and a vanishing longitudinal resistance~\cite{chang2013experimental,chang2015high,mogi2015magnetic}. To enter the field of topological quantum computing, the MTI needs to be proximitized by a superconducting electrode to achieve topological superconductivity. Although induced superconducting correlations in MTIs have already been shown~\cite{uday2023induced} and even an onset of a supercurrent has been measured in an MTI based Josephson junction~\cite{jansen2024josephson}, inducing a hard superconducting gap into this material class remains a challenge. It should be emphasized that in order to realize Majorana fermions and topological superconductivity , the MTI does not necessarily have to be in the quantum anomalous Hall regime~\cite{legendre2024topological}. In fact, by performing tight-binding simulations, it was found that a smaller intrinsic magnetization of the film should result in a larger induced superconducting gap and thus be beneficial for proximitized MTI structures~\cite{burke2024robust}. 
Here we present a systematic in-depth study of the interplay between molecular beam epitaxy (MBE) growth and magneto-transport of MTI thin films. The MTI material investigated in this work is Cr$_x$(Bi$_y$\,Sb$_{1-y}$)$_{2-x}$Te$_3$ and in the frame of this paper multiple sets of thin films were grown. The study focuses mainly on the effect of the growth temperature ($T_{\text{sub}}$) and the Cr concentration  on the crystal quality and the magneto-transport properties of the MTI thin films. We will first explain the growth dynamics and discuss how the material properties change for these different MBE parameters. Then, the effect of substrate temperature on the magneto-transport properties is investigated. Finally, the effect of varying chromium concentration is studied in transport measurements and the results are discussed. Further information on the structural characterization of the films and the methods used for the electrical characterization can be found in the supporting material.

\section{Growth optimization}

To fabricate Cr$_x$(Bi$_y$\,Sb$_{1-y}$)$_{2-x}$Te$_3$ thin films the precise stoichiometric composition of the base alloy (Bi$_y$\,Sb$_{1-y}$)$_2$Te$_3$ which positions the Fermi level inside the bulk gap is crucial. Previous findings~\cite{Jalil:848705,schuffelgen2019selective,jalil2023selective} indicate the optimum stoichiometry to be Bi$_{0.34}$Sb$_{1.66}$Te$_{3}$, having the Bi content at $17\,\%$. The incorporation of Cr into (Bi$_y$\,Sb$_{1-y}$)$_2$Te$_3$ films results in Cr atoms occupying the lattice sites of Bi and Sb. Moreover, Cr, akin to Sb, promotes $p$-type doping in the crystal which requires an increase in Bi content to offset the doping and sustain the Fermi level inside the band gap. The ideal stoichiometry for a quantum anomalous Hall insulator (QAHI), as indicated by Chong et al.~\cite{chong2023electrical} is Bi$_{0.8}$Sb$_{1.2}$Te$_3$, which elevates the Bi content to $40\,\%$. Thus, in the preliminary stage of this work, all individual beam fluxes are set to achieve the target stoichiometry of Cr$_{0.2}$(Bi$_{0.4}$Sb$_{0.6}$)$_{1.8}$Te$_{3}$.  

\subsection{Ideal substrate temperature}

\begin{figure*}[hbtp]
\centering
\includegraphics[width=0.9\textwidth]{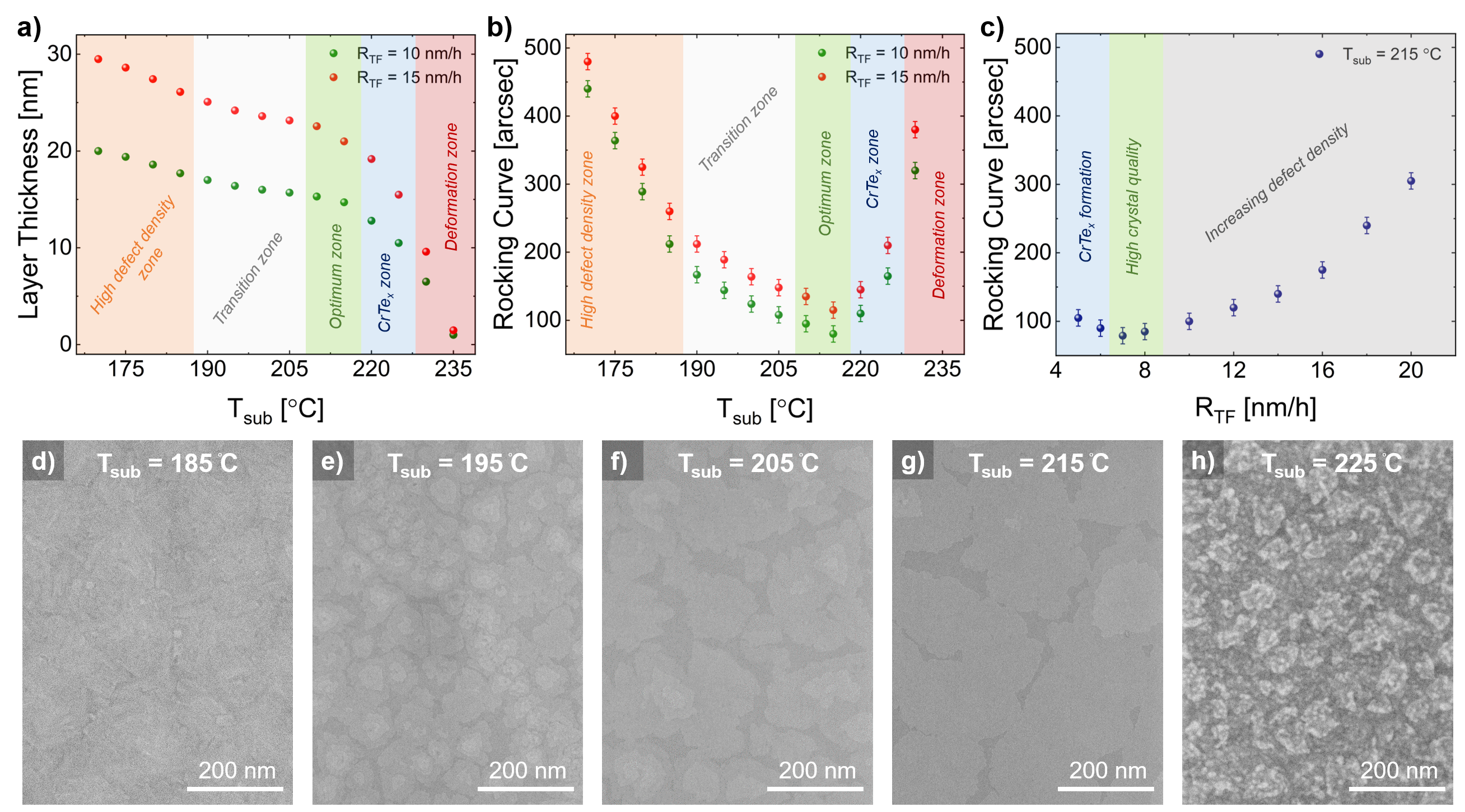}
\caption{Identification of optimal parameters for the growth of the Cr$_x$(Bi$_y$\,Sb$_{1-y}$)$_{2-x}$Te$_3$ thin films via MBE. a) Search for optimal growth temperature $T_{\text{sub}}$ where the layer thicknesses are measured via XRR and b) the crystal quality is assessed by the FWHM value of the rocking curve via XRD measurements. c) The rocking curve FWHM values as a function of growth rate with epilayers prepared at optimal $T_{\text{sub}}$. The green area always denotes the optimum zone of the growth parameters. d) to h) Scanning electron microscopy (SEM) images of Cr$_x$(Bi$_y$\,Sb$_{1-y}$)$_{2-x}$Te$_3$ thin films for increasing substrate temperature. In the aforementioned optimum zone the films are visibly the smoothest.}
\label{Fig_substrate_growth}
\end{figure*}

To determine the ideal growth temperature, the method proposed by Jalil et al.~\cite{jalil2023selective, jalil2023phase} is employed, for which epilayers with thin film growth rates (R$_{\text{TF}}$) of $10$~nm/h and $15$~nm/h are deposited over a temperature range between $170\degree$C to $240\degree$C. Based on the analysis, utilizing x-ray reflectometry (XRR) for thickness measurement and x-ray diffraction (XRD) rocking curve for assessing crystal quality, the entire $T_{\text{sub}}$ range is categorized into several zones, as illustrated in Figs.~\ref{Fig_substrate_growth} a).

As the name indicates in the high defect-density zone, the relatively low $T_{\text{sub}}$ leads to the creation of high density of grains and numerous structural defects, such as domains, rotational twins, and translational shear faults, which in turn results in elevated values of the full width at half maximum (FWHM) of the rocking curves. In the transition zone, the rising $T_{\text{sub}}$ contributes to enhanced crystal quality, resulting in larger grain sizes and subsequently lower FWHM values. The optimum zone yields exceptional crystal quality, featuring the largest grain sizes among all epilayers and significantly fewer structural defects, resulting in the FWHM values below $100$~arcsecond. A further increase in $T_{\text{sub}}$, unlike the growth of (Bi$_y$\,Sb$_{1-y}$)$_2$Te$_3$ alloy~\cite{jalil2023selective}, does not lead to an instant deformation zone. Instead, the existence of an intermediate zone is observed. As the substrate temperature $T_{\text{sub}}$ increases beyond the optimal range, the adsorption-to-desorption ratio of Sb and Bi atoms decreases, which would typically lead to epilayer deformation. However, rather than inducing such a deformation, the strong adsorption of Cr promotes the segregation of Cr atoms from the Cr$_x$(Bi$_y$\,Sb$_{1-y}$)$_{2-x}$Te$_3$ quintuple layer structure. This segregation results in the formation of CrTe$_x$ alloy that leaves behind residual and structurally compromised (Bi$_y$\,Sb$_{1-y}$)$_2$Te$_3$ layers. CrTe$_x$ is known to be an antiferromagnetic material~\cite{PhysRevB.108.094409,PhysRevB.111.035118} and its presence can adversely affect the desired topological properties. Therefore, its formation should be carefully avoided. Further increases in $T_{\text{sub}}$ push the system into the deformation zone, where the adsorption-to-desorption ratio drops sharply, eventually leading to the complete evaporation of the layers. The influence of growth temperature on the surface morphology of the films is also evident in the scanning electron microscopy (SEM) images shown in Figs.~\ref{Fig_substrate_growth} d) to h).

\subsection{Optimal growth rate}
The findings of the previous section conclude that the optimal $T_{\text{sub}}$ is $215\degree$C. While keeping $T_{\text{sub}}$ constant at $215\degree$C, the search for the optimal growth rate is initiated with epilayers being prepared at various R$_{\text{TF}}$ ranging from $5$~nm/h to $20$\,nm/h. Systematically decreasing R$_{\text{TF}}$ from $20$~nm/h to $7$\,nm/h leads to continuous improvement in the crystal quality. However, once R$_{\text{TF}}$ falls below $7$\,nm/h, the low adsorption to desorption ratio of Sb and Bi, combined with the relatively strong adsorption of Cr, results in the formation of CrTe$_x$, similar to as observed in epilayers prepared at higher than optimal $T_{\text{sub}}$ (see Fig.~\ref{Fig_substrate_growth} c)). To prevent this and taking into account the stability of the individual beam fluxes to reach the goal of ultra-low Cr compositions in this study, an R$_{\text{TF}}$ of $8$\,nm/h is selected for all subsequent growths. The corresponding elemental beam fluxes are as follows: Bi $ = 1.6 \times 10^{-8}$\,mbar, Sb $ = 2.2 \times 10^{-8}$\,mbar, and Te $ = 7.6 \times 10^{-7}$\,mbar, keeping the (Bi+Sb):Te ratio at $1$:$20$ to reduce point defects. The Cr beam flux is insufficient for precise measurement and therefore the temperature of the Cr effusion cell ($T_{\text{Cr}}$) is utilized to identify any variations in Cr concentrations.  

\subsection{Chromium concentration}
To obtain consistent thin films with Cr compositions ranging from ultra-low ($<1\,\%$) to relatively high ($15\,\%$), the temperature of the chromium effusion cell ($T_{\text{Cr}}$) is systematically adjusted while keeping the Bi:Sb flux ratio unaltered. Setting $T_{\text{Cr}}$ at $930\degree$C results in no Cr incorporation into the epilayer, as the effective Cr flux is nearly absent. The elemental compositions are measured using Rutherford backscattering spectroscopy (RBS), as outlined in previous research~\cite{Jalil:848705}. Starting at $T_{\text{Cr}} = 945\degree$C, a systematic integration of Cr into the epilayer is observed, with $T_{\text{Cr}} = 950\degree$C resulting in $1\,\%$ and $T_{\text{Cr}} = 1030\degree$C yielding approximately $15\,\%$ Cr contents in the epilayer.

\begin{figure*}[hbtp]
\centering
\includegraphics[width=0.9\textwidth]{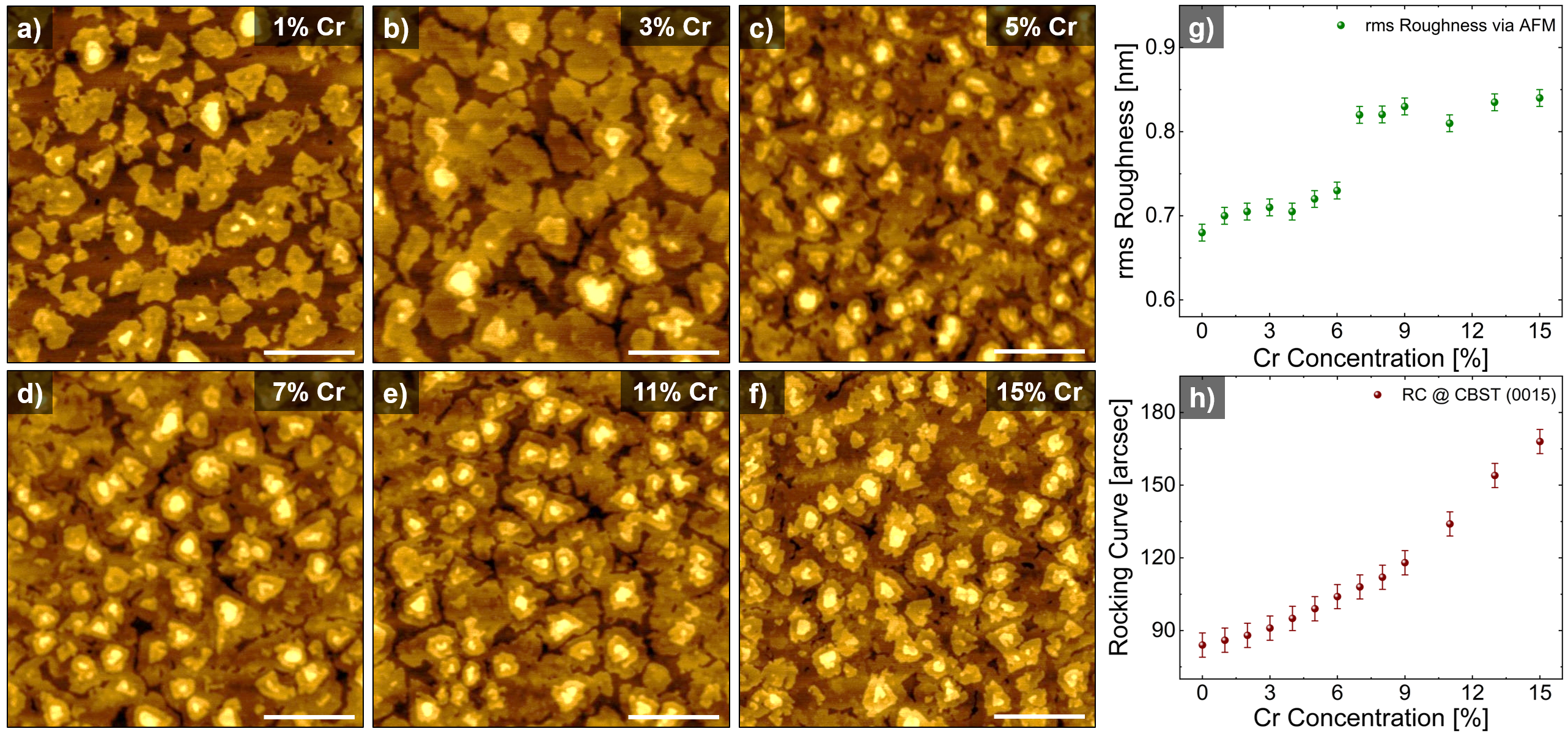}
\caption{the effect of increasing chromium concentration on the surface morphology and crystal structure. a) to f) AFM images of Cr$_x$(Bi$_y$\,Sb$_{1-y}$)$_{2-x}$Te$_3$ thin films with an increasing x from $1$~\% to $15$~\%. g) rms surface roughness of the thin films as a function of Cr concentration and h) The FWHM values of the rocking curve, acquired at the Cr$_x$(Bi$_y$\,Sb$_{1-y}$)$_{2-x}$Te$_3$ (0015) peak in epilayers with increasing Cr contents.}
\label{Fig_chromium_growth}
\end{figure*}

Multiple epilayers containing various Cr concentrations and thicknesses are prepared and systematically investigated to detect any changes in texture, surface roughness, and crystal quality. Observations indicate that in the absence of Cr, the epilayers consistently initiated nucleation at the substrate step edges, resulting in the development of relatively large grains along the edge line. This, in turn, promotes the emergence of unique linear texture. The integration of Cr leads to a significant shift in dynamics, as the strong adsorption of Cr creates additional nucleation sites. This not only alters the film texture but also leads to smaller grains and a relatively diminished crystal quality. The AFM images shown in Figs.~\ref{Fig_chromium_growth} a) to f) reveal a unique change in the surface texture as the Cr concentration increases in the epilayer. Regarding surface roughness, all epilayers exhibit a root mean square (rms) surface roughness of less than $0.85$\,nm. However, a marginal increase was observed in the epilayers with Cr contents beyond $6\,\%$ (cf. Fig.~\ref{Fig_chromium_growth} g)). The rocking curve analysis reveals a steady decline in crystal quality as Cr content increases, mostly attributed to the reduction in grain sizes (cf. Fig.~\ref{Fig_chromium_growth} h)). 

\section{Magneto-transport measurements}
\begin{figure*}[hbtp]
\centering
\includegraphics[width=0.65\textwidth]{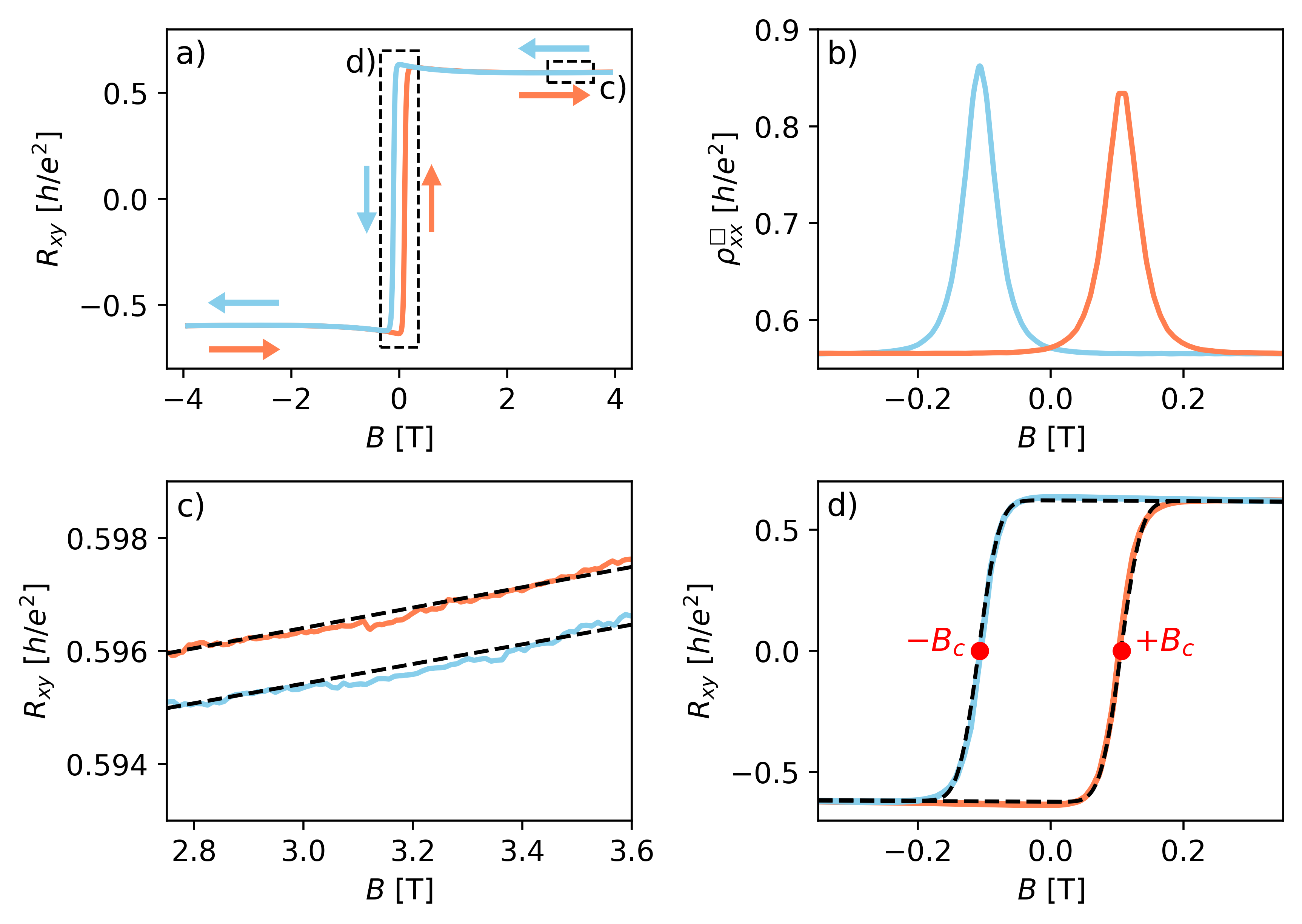}
\caption{Resulting curves from the van der Pauw measurements in an exemplary manner for one of the MTI films with a Cr concentration of 6.5\%. a) Hall resistance $R_{xy}$ in units of $h/e^2$ as a function of magnetic field. The arrows indicate the direction of the magnetic field sweep. The dashed boxes highlight the anomalous part and the regular part of the Hall resistance, respectively. b) Zoom into the longitudinal resistance $\rho_{xx}$ for small magnetic fields. $\rho_{xx}$ exhibits a peak at the positive and negative coercive field. c) Zoom into regular part of the Hall resistance. The dashed line is a linear fit from which the two dimensional charge carrier density $n_{2d}$ is extracted. d) Zoom into the anomalous part of the Hall resistance. The dashed line is a sloped error function fit to the data from which the coercive field $B_c$ and the anomalous Hall resistance $R_{\text{AH}}$ can be determined.}\label{Fig_Rxx_and_Rxy}
\end{figure*}

In order to extract the essential magneto-transport properties of the thin films the Hall resistance $R_{xy}$ and sheet resistivity $\rho^{\square}_{xx}$ are determined for the complete range of magnetic field $B$ by using the van der Pauw technique~\cite{van1958method}. The results are presented in an illustrative manner in Fig.~\ref{Fig_Rxx_and_Rxy} for an MTI film grown at a substrate temperature of $205^\circ$C and a Cr cell temperature of $990^\circ$C, resulting in a Cr concentration of $6.5\%$. The Hall resistance $R_{xy}$ is depicted in Fig.~\ref{Fig_Rxx_and_Rxy}\,a). The sweep direction of the magnetic field is indicated with arrows. It is apparent that the $R_{xy}$ curve consists of a hysteretic part around zero field and shows a linear behaviour on $B$ at larger absolute fields. In Fig.~\ref{Fig_Rxx_and_Rxy}\,b) the sheet resistivity $\rho^\square_{xx}$ is shown for a smaller range of magnetic fields. Two peaks are apparent at the field $B=\pm B_c$, with $B_c$ called the coercive field, where $R_{xy}$ in Fig.~\ref{Fig_Rxx_and_Rxy}\,a) crosses through zero. The two-dimensional charge carrier density $n_{2d}$ of the material is extracted from $R_{xy}$ via a linear fit (dashed line) to data points at higher magnetic fields where the magnetic hysteresis does not affect the shape of the curve anymore (cf. Fig.~\ref{Fig_Rxx_and_Rxy}\,c)). The slight offset of the curves for the different sweep directions of the magnetic field is most likely due to some charging effects. The sheet carrier density is given by $n_{2d}=1/(e R_H)$, 
with $R_H$ being the Hall coefficient which is equal to the slope of the linear fit. The average of $n_{2d}$ for both sweep directions, as well as positive and negative fields, gives the best representation of how many charge carries are present in the material. The mobility $\mu$ can be calculated via $\mu=1/(e n_{2d} \rho^{\square}_{xx}(B=0))$. Here, $\rho^{\square}_{xx}(B=0)$ describes the sheet resistivity at zero magnetic field which can be extracted from the curve shown in Fig.~\ref{Fig_Rxx_and_Rxy}\,b). The anomalous Hall resistance $R_{\text{AH}}$ and the coercive field $B_c$ are extracted from an error function fit (dashed line) to the hysteretic part of the Hall resistance, as illustrated in Fig.~\ref{Fig_Rxx_and_Rxy}\,d)~\cite{zimmermann2024fourier}. The height of the fitted error function is equal to $R_{\text{AH}}$ and the position where the function is zero represents $\pm B_c$, indicated by the red dots.

\subsection{Effect of substrate temperature} 

As mentioned above, the substrate temperature affects the stoichiometry of the material due to changing surface dynamics during MBE growth assuming that all other parameters remain constant. Hence, even small variations in substrate temperature are also expected to change the magneto-transport properties of the film.
The films presented in this section were grown to be within the transition and optimum zone from Fig.~\ref{Fig_substrate_growth} a) with a constant Cr cell temperature of $T_{\text{Cr}}=1015\degree$C.

\begin{figure*}[hbtp]
\centering
\includegraphics[width=0.65\textwidth]{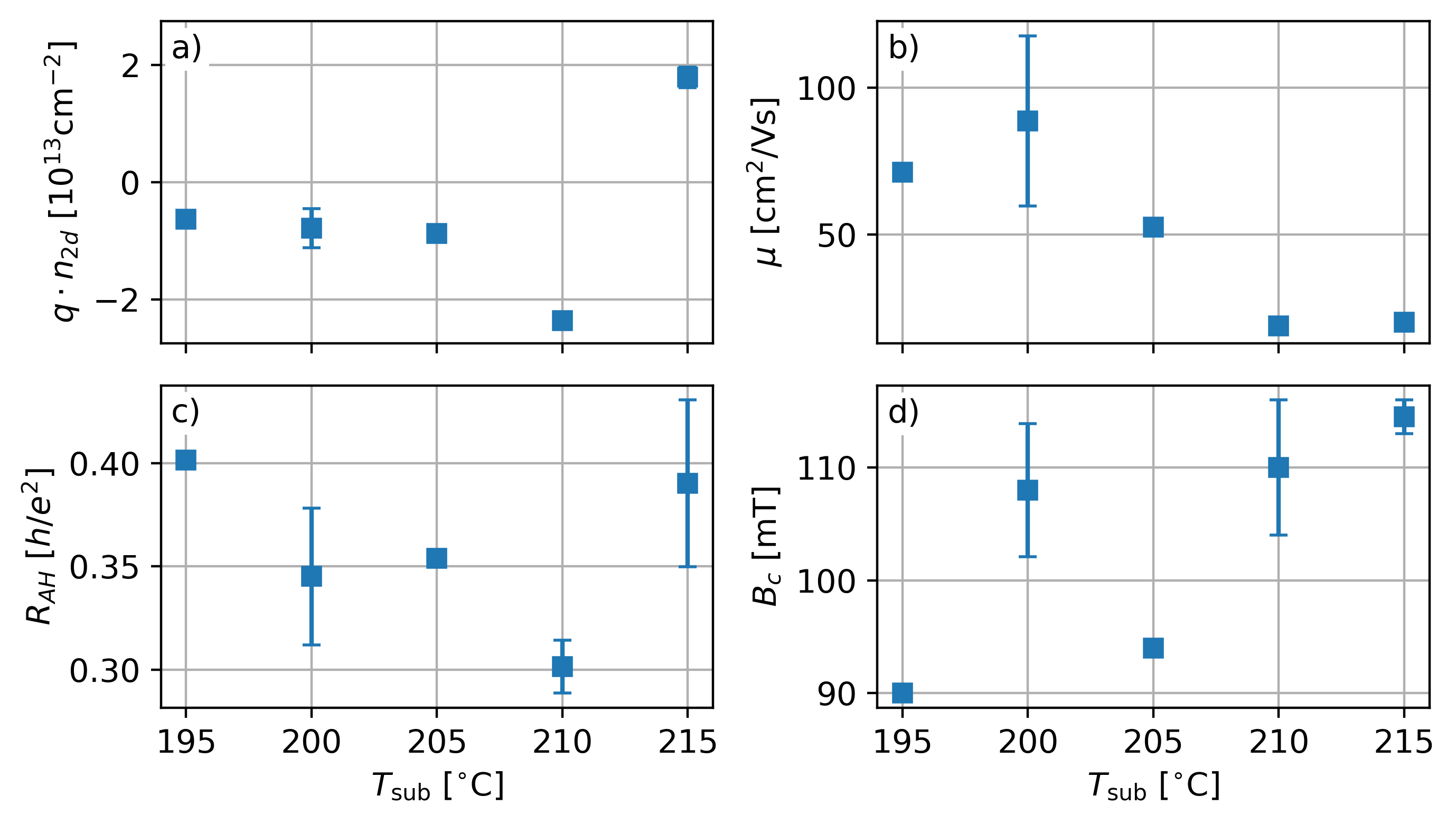}
\caption{Magneto-transport parameters of the thin ($\sim 8$\,nm) Cr$_x$(Bi\,Sb)$_{2-x}$Te$_3$ films as a function of substrate temperature extracted from measurements at $4$\,K. a) Sheet carrier concentration $n_{2d}$ multiplied by the sign of the slope of the Hall signal $q=\pm 1$ for hole and electron dominated transport, respectively. b) Mobility $\mu$ calculated from $n_{2d}$ and $\rho_{xx}(B=0)$. c) Anomalous Hall resistance $R_{\text{AH}}$ and d) coercive field $B_c$ as a function of substrate temperature determined via the error function fit to $R_{xy}$ at small magnetic fields.
Note that in all subplots some data points have error bars while others do not. This is because multiple films were grown for some substrate temperatures, so they were combined into one data point.
} \label{Fig_substrate_transport}
\end{figure*}

Figure~\ref{Fig_substrate_transport} shows the four major transport parameters, i.e. $qn_{2d}$, $\mu$, $R_{\text{AH}}$ and $B_c$, as a function of substrate temperature, extracted from measurements of roughly $20$\,nm thick films at $4$\,K. Here, the $q$ prefactor of the charge carrier density stands for the sign of the Hall slope. This means that for $q=\pm 1$ the dominant charge carrier type contributing to transport are holes or electrons, respectively. As can be seen in Fig.~\ref{Fig_substrate_transport}\,a), for substrate temperatures up to $210^\circ$C, $qn_{2d}$ is negative, i.e. the transport is electron dominated. Up to $205^\circ$C the carrier concentration is relatively low, i.e. below $1 \times 10^{13}$\, cm$^{-2}$. At the highest substrate temperature of $215^\circ$C the transport is hole dominated. This means that the Fermi level shifts from the conduction band to the valence band. The mobility as a function of substrate temperature is shown in Fig.~\ref{Fig_substrate_transport} b). The highest mobility of about $85\,\mathrm{cm}^2/\mathrm{Vs}$ is observed for $T_s=200^\circ$C. For the two highest temperatures the mobility is rather low in the order of $20\,\mathrm{cm}^2/\mathrm{Vs}$, most likely due to the presence of charge puddles~\cite{ito2022cancellation}. As can be seen in Fig.~\ref{Fig_substrate_transport} c), at $4$\,K none of these samples are close to being quantized in Hall resistance,  which is apparent from the value of $R_{\text{AH}}$ always being smaller than $0.5\,h/e^2$. Furthermore, the coercive field $B_c$ shown in Fig.~\ref{Fig_substrate_transport} d) does not really show a systematic dependence on the substrate temperature. It can be summarized that the shape of the anomalous Hall resistance curve, i.e. the height $R_{\text{AH}}$ and the width $B_c$, is not clearly affected by the substrate temperature. In the next section, a more direct way to change the material composition and hence the magnetic properties is presented by means of changing the Cr concentration of the MTI.

\subsection{Effect of chromium concentration}
To systematically vary the Cr concentration in the films, the Cr cell temperature $T_{\text{Cr}}$ was varied between $1015\degree$C and $1022\degree$C for the more strongly doped, both thicker and thinner films, with $T_{\text{sub}}=200\degree$C resulting in Cr concentrations between $11\,\%$ and $15\,\%$. For the weakly doped thin films $T_{\text{Cr}}$ was varied between $950\degree$C to $1010\degree$C at $T_{\text{sub}}=205\degree$C resulting in Cr concentrations between $1\,\%$ and $11.5\,\%$. The substrate temperature for the films with varying Cr contents was chosen to be within the previously determined optimum zone.
\begin{figure*}[hbtp]
\centering
\includegraphics[width=0.85\textwidth]{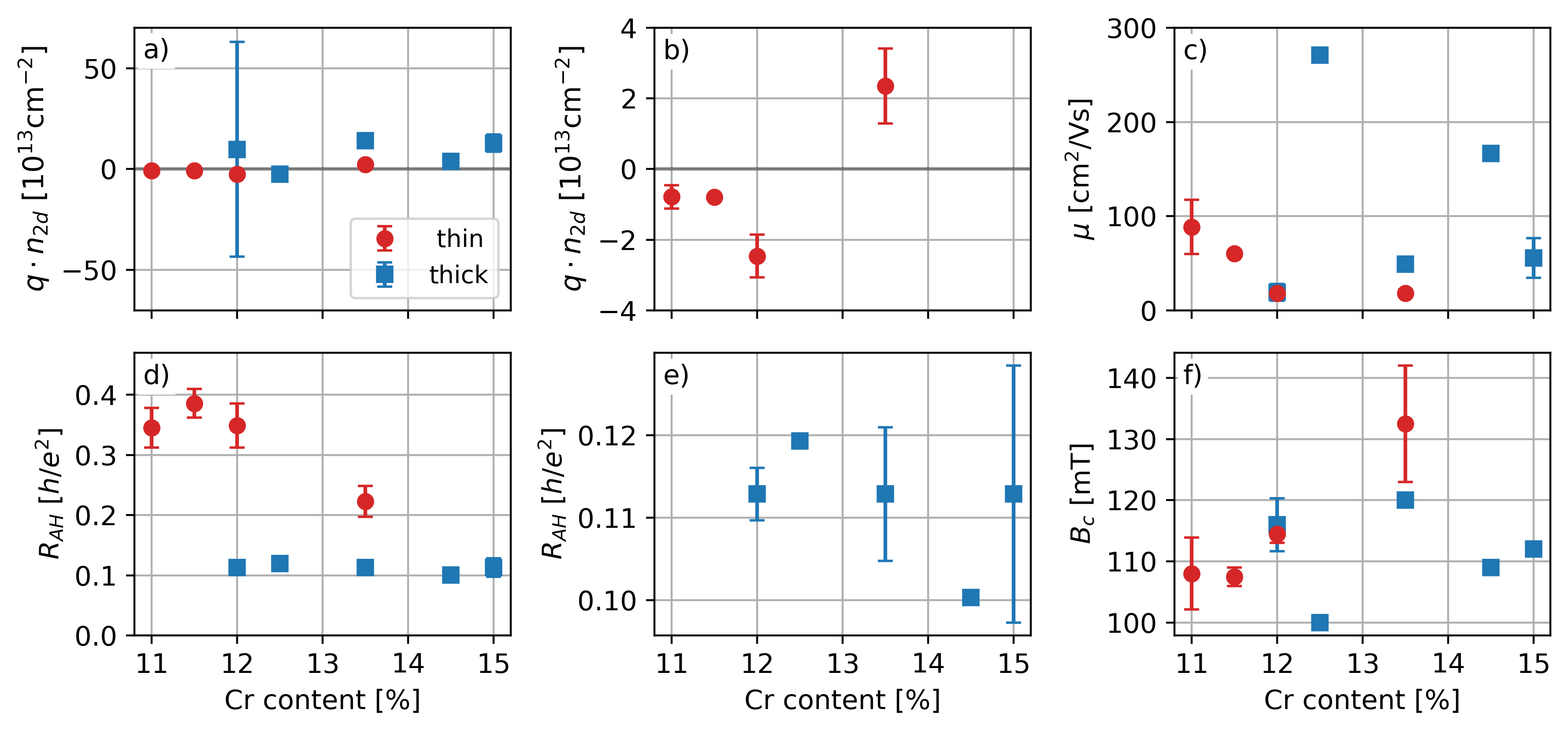}
\caption{Transport parameters of thin ($\sim 8$\,nm) and thick ($\sim 20$\,nm) Cr$_x$(Bi$_y$Sb$_{2-y}$)$_{2-x}$Te$_3$ films measured at $4$\,K as function of Cr-concentration. The values of the thick layers are indicated by blue squares, while the thin samples are represented by red circles. a) Carrier concentration $n_{2d}$ multiplied by the sign of the slope of the Hall signal $q=\pm 1$ for hole and electron dominated transport, respectively. b) Zoom into a smaller range of $n_{2d}$ to highlight the switch from electron to hole transport for the thin films. c) Mobility $\mu$ for thick and thin samples. d) Anomalous Hall resistance $R_{\text{AH}}$ in $h/e^2$ as a function of $x$. The thin samples reach much larger values. e) Zoom into $R_{\text{AH}}$ for the thicker samples. f) Coercive field $B_c$.} \label{Fig_chromium_big_transport}
\end{figure*}
Figure~\ref{Fig_chromium_big_transport} shows once again the four major magneto-transport parameters of the thinner and thicker films wit Cr contents from $11.5\,\%$ to $15\,\%$ extracted from measurements at $4$\,K. The thick films have a thickness of approximately $20$\,nm and the thin films are roughly $8$\,nm thick. As can be seen in Fig.~\ref{Fig_chromium_big_transport} a), in the presented range of Cr concentration, $q n_{2d}$ experiences a sign change for both the thick and the thin samples from initially negative, indicating electron dominated $n$-type transport, to positive values, indicating mainly hole-dominated $p$-type transport. For thicker films, the carrier concentration is an order of magnitude larger than for thinner films, since the surface to bulk ratio is smaller. Therefore, Fig.~\ref{Fig_chromium_big_transport} b) presents a closer look into the dependence of $qn_{2d}$ on the Cr concentration $x$ for the set of thin films. We find a minimal value of $q n_{2d}=-2.48\cdot10^{13}$\,cm$^{-2}$ at $x=12.5\,\%$ for the thick films and $q n_{2d}=-8.5\cdot10^{12}$\,cm$^{-2}$ at $x=11.5\,\%$ for the thin films, respectively. The switch from $n$-type to $p$-type transport for an increasing $x$ is expected since Cr is known to be a $p$-type dopant for the underlying ternary topological insulator (Bi$_x$Sb$_y$)$_2$Te$_3$. This means that increasing the Cr concentration allows tuning the Fermi energy to the exchange gap region of the MTI, where the contribution of the chiral edge state is largest compared to the surface or bulk states.

Normally one would expect the carrier density to become small as $q n_{2d}$ changes sign. However, here it seems more like a divergence before the sign is changed, which is indicated by large error bars for the thick samples with a Cr content of $11.5\,\%$ in Fig.~\ref{Fig_chromium_big_transport} a). The error bars result from the variation of several different samples grown under the same condition. A large error bar indicates that all samples at this Cr concentration have a Fermi level close to the charge neutral point (CNP) of the material. The CNP for gapped bandstructures describes the position of the Fermi energy for which the material is neither n-type nor p-type conductive and hence considered to be charge neutral. Since the potential landscape of the sample can fluctuate due to disorder, when the Fermi energy is located close to the CNP it can cut the surface state conduction and valence band, resulting in $p$-type and $n$-type charge puddles~\cite{Weyrich2016,lippertz2022current}. Since electrons and holes contribute equally to transport in this regime, a vanishing Hall slope is measured, leading to apparent large carrier densities. This effect is also noticeable for the thin samples at $12\%>x>13.5\%$ in Fig.~\ref{Fig_chromium_big_transport} b). For non-magnetic materials, a two-channel model can be applied, which replaces the classical Hall analysis when two different charge carriers are involved~\cite{ashcroft1976solid,ito2022cancellation}. However, due to magnetic hysteresis, such a two-channel model is difficult to apply in our case. 

Regarding the mobility $\mu$ we find that it is approximately inversely proportional to $n_{2d}$ for both thick and thin samples as, shown in Fig.~\ref{Fig_chromium_big_transport} c). It can be seen that the mobility is generally smaller for the thinner samples compared to the thicker ones which 
might be due to the larger relative contribution of surface scattering for the thinner layers. The maximum mobility $\mu=60.7$\,cm$^2$/Vs for the thin films is found at $x=11\,\%$ and the largest mobility for the thick films is given by $\mu=271$\,cm$^2$/Vs at $x=12.5\,\%$.
Another effect of the thickness reduction is seen in $R_{\text{AH}}$, which is presented in Fig.~\ref{Fig_chromium_big_transport} d). Due to the larger surface to bulk ratio, the height of the magnetic hysteresis is a factor of up to three larger for the thin samples compared to the thick ones. This results in the thin samples being up to $40\,\%$ of the quantized value in $R_{\text{AH}}$ at $4$\,K. The thick samples only reach anomalous Hall resistances of around $0.11\,h/e^2$ with no apparent trend as a function of Cr content as shown in Fig.~\ref{Fig_chromium_big_transport} e). Finally, the coercive field $B_c$ shown in Fig.~\ref{Fig_chromium_big_transport} f) is not much affected by the thickness reduction, although for the thin samples a systematic increase with Cr concentration is observed. This is due to the presence of more ferromagnetic Cr atoms in the film, which leads to a larger exchange gap due to a stronger magnetization of the film~\cite{Lu2011competition,chen2010massive}.Therefore, an indirect indicator of the size of the magnetic exchange gap that is accessible from transport measurements is the coercive field $B_c$.

\begin{figure}[hbtp]
\centering
\includegraphics[width=0.49\textwidth]{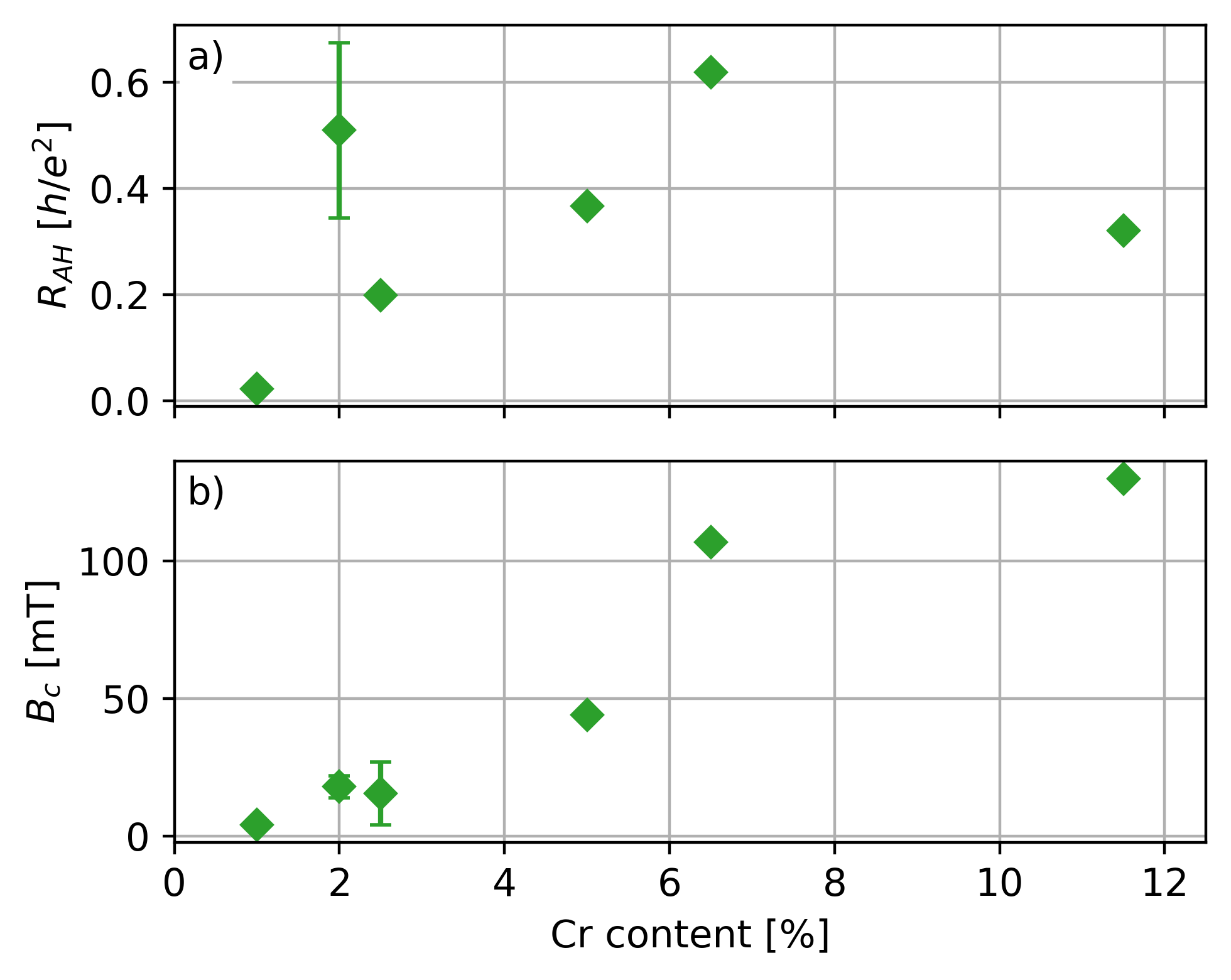}
\caption{a) Anomalous Hall resistance $R_{\text{AH}}$ and b) coercive field $B_c$ determined at a temperature of 1.2\,K as a function of the Cr concentration $x$ in a range of $1\,\%$ to $11.5\,\%$.}
\label{Fig_chromium_small_transport}
\end{figure}

Our findings, as well as previous studies indicate that one should be able to fine tune the coercive field to a small desired value by fine tuning the Cr concentrations in a smaller regime~\cite{kou2013interplay}. Therefore, the fourth set of samples with smaller overall Cr concentrations was grown and investigated by transport measurements. Figure~\ref{Fig_chromium_small_transport} a) shows $R_{\text{AH}}$ for a range of $1\,\%>x>11.5\,\%$. Although this data set was not optimized with respect to charge neutrality, $R_{\text{AH}}$ reaches more than $0.6\,h/e^2$ for two different Cr concentrations. One of the concentrations is $6.5$\,\% which is comparable to Cr concentrations in quantized samples \cite{chang2013experimental}. However, the other high value of $R_{\text{AH}}$ averaging around $0.5\,h/e^2$ is found at a very low Cr concentration of only 2\,\%. The coercive field $B_c$ shown in Fig.~\ref{Fig_chromium_small_transport} b) only measures $18$\,mT at this Cr concentration. It is further apparent that it is indeed possible to fine-tune $B_c$ by varying the Cr concentration. At the lowest value of $x$, the films show almost no magnetic properties anymore. Subsequently, the coercive field increases almost linearly with Cr concentration. A crucial finding is that it is possible to obtain low values of $B_c$, and thus small magnetic exchange gaps, in a reproducible fashion. As mentioned before, films with a smaller $B_c$ and hence a smaller magnetization, resulting in a smaller magnetic exchange gap, are theoretically predicted to be more beneficial for inducing superconductivity into the MTI.

\section{Conclusion}

Magnetic topological insulators are predicted to have interesting applications in quantum computation. Controlling the parameters of the molecular beam epitaxy during growth of the MTI thin films allows to reach exceptional crystal quality and low surface roughness in combination with large grain sizes. Furthermore the MBE enables precise control over the material composition which can be used to alter the magneto-transport properties of the thin films. This allows to tune the MTI thin films into a regime where strong induced superconducting correlations are expected when combining them with conventional superconductors. 

\section*{Acknowledgments}

We thank Herbert Kertz for technical assistance. This work is supported by the QuantERA grant MAGMA and by the Deutsche Forschungsgemeinschaft (DFG, German Research Foundation) under Grant No. 491798118. In addition, we received funding from DFG under Germany’s Excellence Strategy - Cluster of Excellence Matter and Light for Quantum Computing (ML4Q) EXC 2004/1 – 390534769 as well as financial support by the Bavarian Ministry of Economic Affairs, Regional Development and Energy within Bavaria’s High-Tech Agenda Project "Bausteine für das Quantencomputing auf Basis topologischer Materialien mit experimentellen und theoretischen Ansätzen" (grant no. 07 02/686 58/1/21 1/22 2/23).

\section*{Data Availability}

The data that support the findings of this article are openly
available \cite{Karthein-data2025}.

\cleardoublepage
\widetext

\titleformat{\section}[hang]{\bfseries}{\MakeUppercase{Supplemental Note} \thesection:\ }{0pt}{\MakeUppercase}
\setcounter{section}{0}
\setcounter{equation}{0}
\setcounter{figure}{0}
\setcounter{table}{0}
\setcounter{page}{1}

\renewcommand{\thesection}{\arabic{section}}
\renewcommand{\theequation}{S\arabic{equation}}
\renewcommand{\thefigure}{S\arabic{figure}}
\renewcommand{\figurename}{Supplemental Figure}
\renewcommand{\tablename}{Supplemental Table}
\renewcommand{\bibnumfmt}[1]{[S#1]}
\renewcommand{\citenumfont}[1]{S#1}

\begin{center}
\textbf{\large Supporting Information: Controlling the magneto-transport properties of magnetic topological insulator thin films from Cr$_x$(Bi$_y$\,Sb$_{1-y}$)$_{2-x}$Te$_3$ via molecular beam epitaxy}
\end{center}

\section{Structural characterization}
\begin{figure*}[hbtp]
\centering
\includegraphics[width=0.9\textwidth]{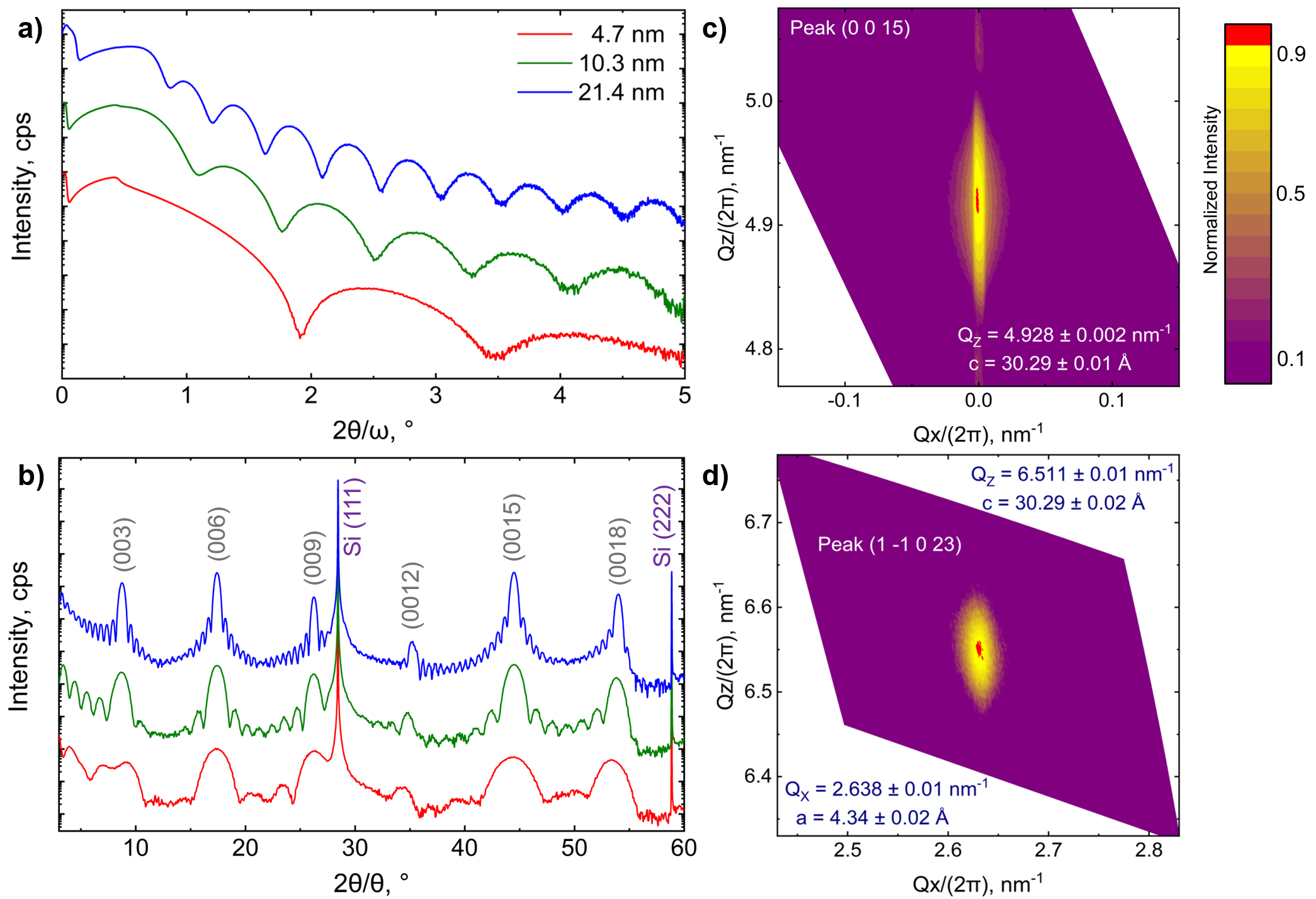}
\caption{a) XRR and b) XRD patterns for approximately $5$, $10$, and $20$\,nm thick epilayers of CrBST with $5\,\%$ Cr content. c) Symmetric reciprocal space map across the CrBST ($0$ $0$ $15$) peak confirming the absence of any tilted domain. d) Corresponding reciprocal space map for the ($1$ $-1$ $0$ $23$) peak.}
\label{Fig_thickness_dep}
\end{figure*}

\noindent All prepared Cr$_x$(Bi$_y$\,Sb$_{1-y}$)$_{2-x}$Te$_3$ (CrBST) epilayers undergo thorough structural investigation via x-ray diffraction (XRD). The detailed information about structural characterization can be found here~\cite{jalil2023phase_sup}. The full width at half maximum (FWHM) of the rocking curve, obtained at the CrBST (0 0 15) peak, serves as the figure of merit during the growth optimization process. The x-ray reflectometry (XRR) investigations have yielded insights into the variations in thickness and roughness of the epilayers. The phase orientation of the grown epilayers is validated through XRD $\Theta⁄2\Theta$ scans, which affirm the single crystalline nature of the CrBST epilayers. The observed smooth surfaces contribute to the appearance of Laue oscillations extending up to 60$\degree$ in the diffraction pattern. Figures~\ref{Fig_thickness_dep} a) and b) illustrate the XRR and XRD patterns for approximately $5$, $10$, and $20$\,nm thick epilayers of CrBST with $5\,\%$ Cr content. A meticulous analysis of the diffraction intensity reveals the presence of Sb with the detection of ($0$ $0$ $9$) peak, but its reduced strength relative to the ($0$ $0$ $6$) peak indicates significant Bi content in the epilayer confirming the aimed Cr$_x$(Bi$_{0.8}$Sb$_{1.2}$)$_{2-x}$Te$_3$ stoichiometry.

\noindent Eventually, the lattice parameters are evaluated through the analysis of both the symmetric and asymmetric reciprocal space maps (RSMs). Supplemental Fig.~\ref{Fig_thickness_dep} c) represents a symmetric map across the CrBST ($0$ $0$ $15$) peak confirming the absence of any tilted domain. The obtained value indicates $Q_Z = 2 \pi (4.928) \pm 0.002$\,nm$^{-1}$, leading to an out-of-plane lattice parameter $c = 30.29 \pm 0.01$\,\AA. This acquired value is in good agreement with previously reported values in the literature~\cite{chong2023electrical_sup,qiu2023manipulating_sup}. It is noteworthy that the incorporation of Cr into the lattice leads to a slight reduction in the out-of-plane lattice, attributed to the smaller size of Cr atoms. The obtained $Q_X$ value is zero, which means that no information regarding the in-plane lattice can be derived from the symmetric RSM. Based on diffraction characteristics of CrBST, the asymmetric ($1$ $-1$ $0$ $23$) peak is selected to probe the in-plane structure, allowing for the extraction of lattice information. Supplemental Fig.~\ref{Fig_thickness_dep} d) illustrates the corresponding RSM. Given that the in-plane lattices are identical due to the trigonal crystal structure, a single asymmetric peak is adequate for assessing the lattice information.
Following the successful epitaxy and structural characterization of the CrBST epilayers, the subsequent section addresses the magneto-transport properties of the grown epilayers.

\section{Experimental}

\noindent All Cr$_x$(Bi$_y$\,Sb$_{1-y}$)$_{2-x}$Te$_3$ films were grown via molecular beam epitaxy (MBE). When investigating the effect of substrate temperature all other parameters, for example the cell temperatures of all the elements, were kept constant. The films with varying substrate temperature from $T_{\text{sub}}=195\degree$C to $T_{\text{sub}}=215\degree$C had cell temperatures of $T_{\text{Cr}}=1015\degree$C, $T_{\text{Bi}}=430\degree$C, $T_{\text{Sb}}=360\degree$C, and $T_{\text{Te}}=420\degree$C. To achieve a varying Cr concentration in the films, the Cr cell temperature $T_{\text{Cr}}$ was varied between $1015\degree$C and $1022\degree$C for the more strongly doped, both thicker and thinner films, with $T_{\text{sub}}=200\degree$C and the remaining cell temperatures being the same as before. For the weakly doped thin films $T_{\text{Cr}}$ was varied between $950\degree$C to $1010\degree$C at $T_{\text{sub}}=205\degree$C. The substrate temperature for the films with varying Cr contents was chosen with respect to the best film quality which was determined by scanning electron microscopy (SEM), XRD and atomic force microscopy (AFM). The stoichiometry of the films, including the exact Cr content, was determined via Rutherford backscattering (RBS).  

\noindent For transport experiments, the sample is loaded into either a liquid helium-based flow cryostat or a $^4$He variable temperature insert (VTI) cryostat. The flow cryostat operates at a base temperature of 4\,K with magnetic fields up to 1\,T. The VTI cryostat can reach 1.2\,K and magnetic fields up to 14\,T. At each base temperature a van der Pauw measurement is performed using two lock-in amplifiers. In order to measure the longitudinal and transverse voltage simultaneously, two alternating currents (AC) of different frequencies are driven through the film by one lock-in amplifier each. The AC current amplitude is set by a fixed resistor in series with the AC voltage generated by the lock-in amplifiers. 

\section{Van der Pauw Measurements}

\noindent In order to ensure a quick feedback on the film properties the van der Pauw method is employed to obtain the relevant transport parameters~\cite{van1958method_sup}. The MTI films are processed immediately after taken out of the MBE chamber. To use the van der Pauw scheme, four metal contacts are sputtered onto each corner of the square-shaped chip using a hard mask. Each chip is then placed in an appropriate chip holder and bonded. The advantage of our approach is that no additional patterning of the thin film is required and the sample can be loaded into the cryogenic measurement setup about one hour after leaving the MBE chamber. This avoids prolonged exposure of the MTI thin films to air, which would lead to surface degradation by oxidation if the film is not protected by a passivation layer, as in our case. 
\begin{figure}[hbtp]
\centering
\includegraphics[width=0.75\textwidth]{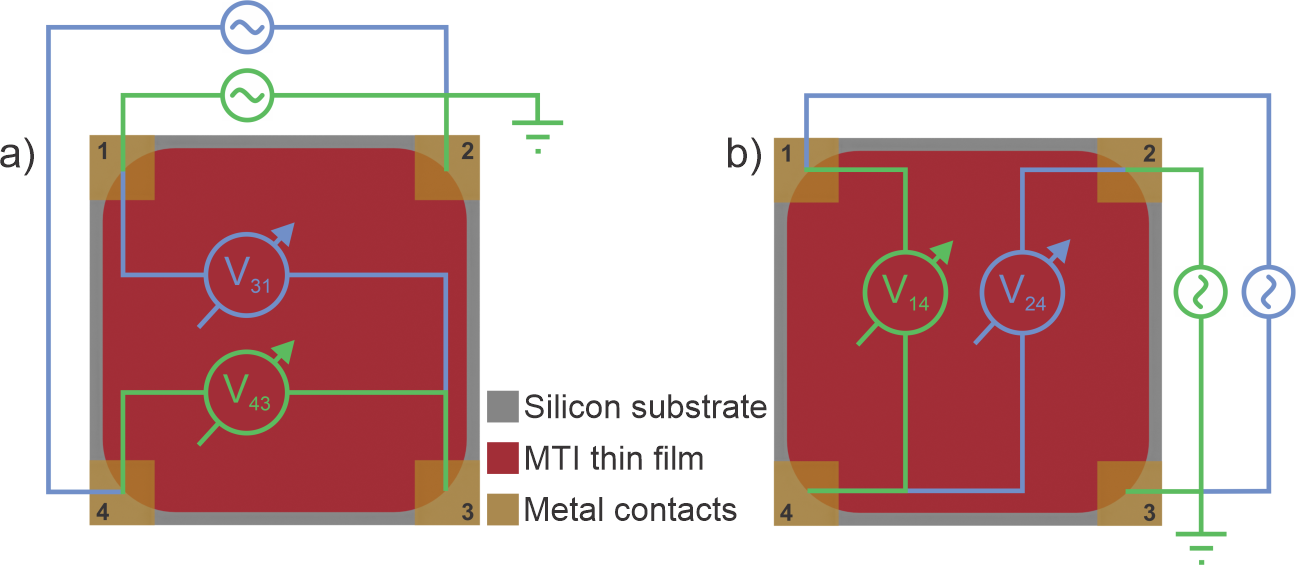}
\caption{Schematic of a van der Pauw measurement. a) First measurement to obtain a set of Hall voltage $V_{31}$ and longitudinal voltage $V_{43}$. b) Measurement with rotated contacting compared to a). This results in a second set of Hall voltage $V_{24}$ and a longitudinal $V_{14}$ voltage.} \label{Fig_vdP_scheme}
\end{figure}
\noindent A simplified measurement scheme is shown in Supplemental Fig.~\ref{Fig_vdP_scheme}. For the first set of measurement (see Supplemental Fig.~\ref{Fig_vdP_scheme}\,a)), two AC-currents are driven through the film at different frequencies using one lock-in amplifier each. This allows to simultaneously measure two voltages and hence cut the measurement time in half. The AC-voltage generated by the lock-in is converted into an AC-current via a fixed resistor. A longitudinal voltage $V_{43}$ is measured by applying an AC-current along the upper edge of the film and measuring the voltage parallel to it at the lower edge (green). A Hall voltage $V_{31}$ is measured by driving another AC-current across the film and measuring the voltage perpendicular to it (blue). To perform the second measurement all contacts are permuted by one in a clockwise manner (see Supplemental Fig.~\ref{Fig_vdP_scheme}\,b)). Similar to the first measurement, a second set of longitudinal $V_{14}$ and Hall $V_{24}$ voltages can be acquired. The sheet resistivity $\rho^{\square}_{xx}$ of the film can be extracted from the longitudinal  voltages by numerically solving the van der Pauw equation
\begin{equation}
    \exp{\Big(-\pi\frac{R_{43}}{\rho^{\square}_{xx}}\Big)}+\exp{\Big(-\pi\frac{R_{14}}{\rho^{\square}_{xx}}\Big)}=1,
\end{equation}
where $R_{ij}=V_{ij}/I$ with $I$ being the AC-current used to measure the longitudinal signals, which is measured with the lock-in amplifier. The Hall resistance $R_{xy}$ is calculated via averaging $R_{31}$ and $R_{24}$ that are calculated with the AC-current used for the Hall signals:
\begin{equation}
    R_{xy}=\frac{R_{31}+R_{24}}{2}.
\end{equation}
In order to extract the essential magneto-transport properties of the thin films $\rho^{\square}_{xx}$ and $R_{xy}$ are determined for the complete range of magnetic field.

\section{Samples}

\noindent The relevant growth and transport parameters of all samples presented in the main text can be found in Supplemental tables~\ref{tab_substrate}-\ref{tab_Crsmall}

\begin{table}[]
    \centering
    \begin{tabular}
    {|c|c|c|c|c|c|c|c|c|c|c|c|}
    \hline
     sample & \makecell{$T_{\text{sub}}$ \\ $[\degree C]$} & \makecell{$T_{\text{Cr}}$ \\ $[\degree C]$} & $x$ & $y$ & $2-y$ & \makecell{d \\ $[$nm$]$} & \makecell{$q\cdot n_{2d}$ \\ $[10^{12}$cm$^{-2}]$} & \makecell{$\mu$ \\ $[$cm$^2$/Vs$]$} & \makecell{$R_{\text{AH}}$ \\ $[h/e^2]$} & \makecell{$B_{c}$ \\ $[$mT$]$}& \makecell{$T$ \\ $[$K$]$}\\
     \hline
     1&195&1015&0.27&0.68&1.05&8.8&-6.33&322.5&10.37&90&4\\
     \hline
     2&200&1015&0.18&0.64&1.18&9.6&-12.5&49.3&7.72&116&4\\
     \hline
     3&200&1015&0.22&0.69&1.08&-&-5.13&118.1&9.7&106&4\\
     \hline
     4&200&1015&-&-&-&-&-5.19&98.4&9.31&102&4\\
     \hline
     5&205&1015&0.27&0.72&1.01&7.9&-8.70&237.2&9.14&94&4\\
     \hline
     6&210&1015&0.26&0.76&0.99&8.0&-22.6&18.1&8.11&104&4\\
     \hline
     7&210&1015&-&-&-&-&-24.6&20.2&7.45&116&4\\
     \hline
     8&215&1015&0.15&0.79&1.06&7.4&19.7&20.1&9.03&116&4\\
     \hline
     9&215&1015&-&-&-&-&16.1&20.5&11.12&113&4\\
     \hline
    \end{tabular}
    \caption{Growth and transport parameters of all samples within the growth temperature study presented in Supplemental Fig.~4 of the main text.}
    \label{tab_substrate}
\end{table}

\begin{table}[]
    \centering
    \begin{tabular}
    {|c|c|c|c|c|c|c|c|c|c|c|c|}
    \hline
     sample & \makecell{$T_{\text{sub}}$ \\ $[\degree C]$} & \makecell{$T_{\text{Cr}}$ \\ $[\degree C]$} & $x$ & $y$ & $2-y$ & \makecell{d \\ $[$nm$]$} & \makecell{$q\cdot n_{2d}$ \\ $[10^{12}$cm$^{-2}]$} & \makecell{$\mu$ \\ $[$cm$^2$/Vs$]$} & \makecell{$R_{\text{AH}}$ \\ $[h/e^2]$} & \makecell{$B_{c}$ \\ $[$mT$]$}& \makecell{$T$ \\ $[$K$]$}\\
     \hline
     10&200&1018&0.23&0.66&1.11&8.7&-8.5&60.7&9.24&106&4\\
     \hline
     11&200&1018&-&-&-&-&-7.45&5-9.6&10.57&109&4\\
     \hline
     12&200&1020&0.24&0.65&1.11&9.4&-30.70&15.4&8.06&113&4\\
     \hline
     13&200&1020&-&-&-&-&-18.60&20.9&9.95&116&4\\
     \hline
     14&200&1022&0.27&0.61&1.13&9.8&34.10&14.9&6.41&123&4\\
     \hline
     15&200&1022&-&-&-&10&12.90&21.9&5.09&142&4\\
     \hline
     16&205&1015&0.25&0.76&0.99&15&-24.8&271&3.08&100&4\\
     \hline
     17&205&1018&-&-&-&-&847.0&8.2&2.82&112&4\\
     \hline
     18&205&1018&-&-&-&-&-208.0&30&3.02&114&4\\
     \hline
     19&205&1018&-&-&-&-&-345.0&19.7&2.9&122&4\\
     \hline
     20&205&1020&0.27&0.70&1.03&-&149.0&45.5&2.63&119&4\\
     \hline
     21&205&1020&-&-&-&-&131.0&53&2.83&121&4\\
     \hline
     22&205&1022&-&-&-&15.4&171.0&34.8&2.40&111&4\\
     \hline
     23&205&1022&0.30&0.66&1.03&18.1&86.3&76.8&2.67&113&4\\
     \hline
     24&205&1025&0.29&0.65&1.07&17.2&38.3&166.4&2.59&109&4\\
     \hline
    \end{tabular}
    \caption{Growth and transport parameters of all samples within the high Cr concentration study presented in Supplemental Fig.~5 of the main text.}
    \label{tab_Crbig}
\end{table}

\begin{table}[]
    \centering
    \begin{tabular}
    {|c|c|c|c|c|c|c|c|c|c|c|c|}
    \hline
     sample & \makecell{$T_{\text{sub}}$ \\ $[\degree C]$} & \makecell{$T_{\text{Cr}}$ \\ $[\degree C]$} & $x$ & $y$ & $2-y$ & \makecell{d \\ $[$nm$]$} & \makecell{$q\cdot n_{2d}$ \\ $[10^{12}$cm$^{-2}]$} & \makecell{$\mu$ \\ $[$cm$^2$/Vs$]$} & \makecell{$R_{\text{AH}}$ \\ $[h/e^2]$} & \makecell{$B_{c}$ \\ $[$mT$]$}& \makecell{$T$ \\ $[$K$]$}\\
     \hline
     25&205&930&0.02&0.77&1.21&8.6&-1.47&558&0.57&4&1.2\\
     \hline
     26&205&950&0.05&0.68&1.27&9.1&1.53&204&4.89&4&1.2\\
     \hline
     27&205&955&0.04&0.72&1.24&8.6&1.19&212.6&8.87&14&1.2\\
     \hline
     28&205&960&0.05&0.71&1.23&8.5&-6.34&53.7&5.36&27&1.2\\
     \hline
     20&205&965&0.04&0.73&1.23&-&1.77&185.2&17.42&22&1.2\\
     \hline
     30&205&970&0.10&0.75&1.16&8.6&18.8&16.7&9.45&44&1.2\\
     \hline
     31&205&990&0.13&0.70&1.17&8.9&9.77&43.6&15.97&107&1.2\\
     \hline
     32&205&110&0.23&0.67&1.1&9.3&24.3&24.3&8.26&130&1.2\\
     \hline
    \end{tabular}
    \caption{Growth and transport parameters of all samples within the low Cr concentration study presented in Supplemental Fig.~6 of the main text.}
    \label{tab_Crsmall}
\end{table}

\end{document}